\documentstyle[epsf,epsfig,12pt]{article}
\begin{document}

\def\grau{$^\circ$}



\begin{titlepage}

\pagenumbering{arabic}
\vspace*{-1.5cm}

\center{\sc EUROPEAN LABORATORY FOR PARTICLE PHYSICS\vspace*{1.5cm}} 
\begin{tabular*}{15.cm}{l@{\extracolsep{\fill}}r}
& CERN-OPEN 98-026
\\
&
  6 September, 1998
\\
&\\ 
\end{tabular*}
\vspace*{2.cm}

\begin{center}
\Large
{\bf
The Estimation of the  Effective Centre of Mass
Energy in q\boldmath{$\overline{\mbox{q}}\gamma$} Events from DELPHI
} \\
\vspace*{2.cm}

\normalsize {
   {\bf P.~Abreu\footnote{LIP/FCUL/IST, Lisboa, Portugal},
        A.~De~Angelis\footnote{CERN-CH 1211, Geneva, Switzerland},
        G.~Della Ricca\footnote{INFN Trieste and Universit\`a di 
        Trieste, Italy},
        D.~Fassouliotis\footnote{University of Athens, Greece},
        A.~Grefrath\footnote{University of Wuppertal, Germany},
        N.~Kjaer\footnote{NIKHEF, Amsterdam, The Netherlands},
        R.P.~Henriques$^1$,
        M.~Mulders$^6$,
        M.~Pimenta$^1$,
        L.~Vitale$^3$
}}
\end{center}

\vspace{\fill}

\begin{abstract}
\noindent
The photon radiation in the initial state lowers the energy available for the
e$^+$e$^-$ collisions; this effect is particularly important at 
LEP2 energies (above the mass of the Z boson). 
Being aligned to the beam direction, such 
initial state
radiation is mostly undetected.
This article describes the procedure used by the DELPHI experiment at LEP to
estimate the effective centre-of-mass energy in hadronic events collected
at energies above the Z peak. Typical resolutions 
ranging from  2 to 3 GeV on the effective center-of-mass energy 
are achieved, depending on the event topology.\\
\noindent{\em PACS codes: 07.05.K, 13.10.+q, 13.65.+i.
Keywords: Data Analysis, Photon Radiation, Initial State Radiation.}


\end{abstract}
\vspace{\fill}

%

\vspace{\fill}
\end{titlepage}

\newpage

\setcounter{page}{1}



\section{Introduction}  \indent

The increase in the LEP beam energy above the Z in 1995 opened a new window of 
higher centre-of-mass (c.m.) energies to study e$^+$e$^-$ annihilations.
The Initial State Radiation reduces the energy available
for the physical processes. 

\begin{figure}
\begin{center}
\begin{minipage}{0.8\linewidth}
\mbox{\epsfig{file=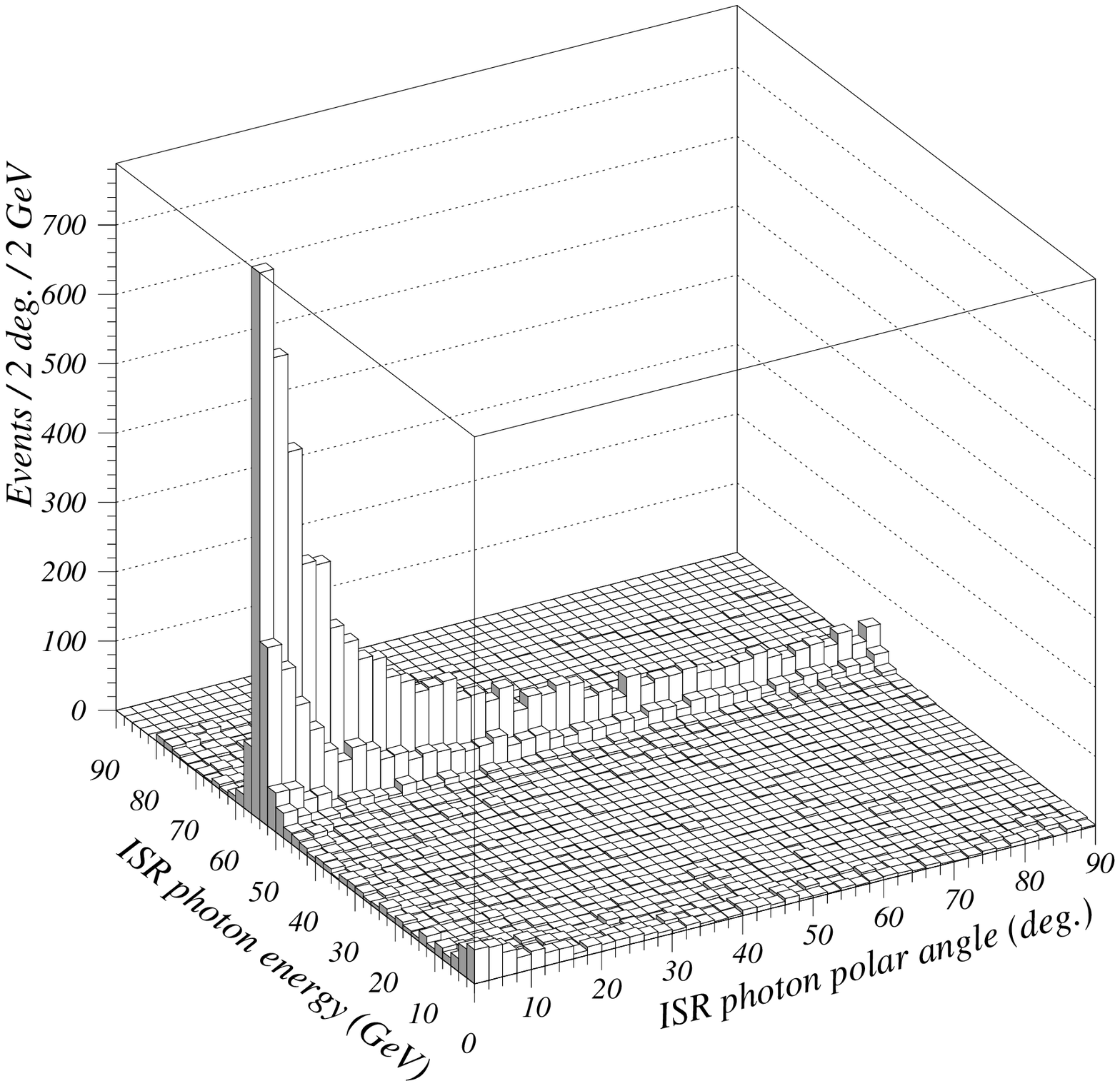,width=\linewidth}}
\caption{\label{sprfig1}
Energy as a function of polar angle for 
ISR photons at c.m. energy of 161 GeV. Photons below 2 degrees are
not plotted.}
\end{minipage}
\end{center}
\end{figure}

The photons from Initial State Radiation (hereafter called ISR photons)
tend to move along the beam ($z$)
direction; thus, they are mostly undetected.
Figure~\ref{sprfig1}
shows the energy versus  polar angle distribution of the ISR photon(s)
in simulated hadronic events at a c.m. energy of
161 GeV. The initial q$\overline{\mathrm{q}}$ state was generated
using the PYTHIA 5.4~\cite{pythia}
Parton Shower Monte Carlo.
The polar angle, $\theta$, 
peaks at values close to zero, and the
photon energies cluster at $E_{\gamma R} = (s-M_Z^2)/ (2\sqrt{s})$, where
$s=(2E_b)^2$, $E_b$ is the beam energy, and $M_Z$ is the Z mass.
For events with such a hard ISR, the
energy in the c.m. of the e$^+$e$^-$ collision is
reduced to the Z mass (``Z radiative return'' events).

This note explains the methods 
used by the DELPHI experiment at LEP to compute the effective centre of mass 
energy, $\sqrt{s^\prime}$, in e$^+$e$^-$ annihilations into hadrons, 
by measuring or estimating the energy and momentum of ISR photon(s).
Two algorithms are used: 
\begin{itemize}
\item the first one, based on the jet directions only, is more robust; 
\item the second, based on a constrained fit, is more precise.
\end{itemize}

The procedures described in this note were used in many papers from 
DELPHI (see for example~\cite{impressivelist,Wmass}).
The DELPHI detector is described in~\cite{delphi}, and its performance 
in~\cite{perfo}.

\section{The methods}\indent

One looks first for detected ISR photons. 
Showers unassociated to charged particles
are searched for in the calorimeters, and they 
are considered as candidate ISR photons, if they
fulfill the following criteria:
\begin{itemize}
\item energy deposition larger than 10~GeV;
\item isolation angle larger than 0.3 radians with respect to any charged 
particle with momentum larger than 1~GeV;
\item when the shower energy is smaller than 70~\% of the photon
energy corresponding to a radiative return to the Z,
isolation angle with respect to any charged particle
(with momentum larger than 1~GeV) larger than 0.6~radians 
and smaller than 1.5~radians (1.8~radians if the shower is contained in 
the small angle calorimeter, the STIC).
\end{itemize}

 Showers are grouped together, if they are closer than 5
degrees in space. The association is done by adding the
energies and assuming as direction the direction of the
vector sum of the momenta.

\subsection{Method 1: determination of the energy of ISR photon(s)
 from the direction of jets} \indent

The first method (called SPRIME inside the collaboration)
uses only the direction of jets and of the detected photon(s) - if any. 

All the particles in the event, except the one(s) considered as ISR photon(s)
using the criteria described above, 
are forced to be clustered in two jets, using the
DURHAM algorithm~\cite{durham}. The particles are assumed to be massless.

The method applies energy and momentum conservation to estimate the
energy of possible ISR photons 
lost along a direction close to the beam, 
after having considered the candidate one(s)
directly detected in the electromagnetic calorimeters.

If no ISR photons are detected, 
an event plane containing the beam $(z)$ axis and
maximizing the sum of the
moduli of the projections of the jet momenta is determined.
A hypothetical photon is assumed to
be along $z$, in the positive or negative direction, respectively if the sum of
the jet polar angles is 
larger or smaller than 180 degrees. The 
directions of the two jets,
projected onto the event plane,
are used to constrain the photon energy.

If only one ISR photon is measured in the detector and the event is planar,
i.e., if the sum of the internal angles between the jets and the photon exceeds
345 degrees, the photon energy is calculated by energy-momentum conservation
in the event plane
containing the photon direction and
maximizing the sum of the
moduli of the projections of the jet momenta.
For aplanar events at least one particle has been lost. The algorithm assumes
that at another photon has been radiated inside the beam pipe and makes
use of the energies of the jets and of the measured ISR photon to constrain
the hypothetical photon energy.

If more than one ISR photon is measured, energy-momentum conservation
is applied in the event plane
to determine the energy of
an unseen photon along $z$ and the momenta of the two jets, 
using as an input the direction of the jets and the energies and the
directions of the detected photons.

\subsubsection{Results}

\begin{figure}
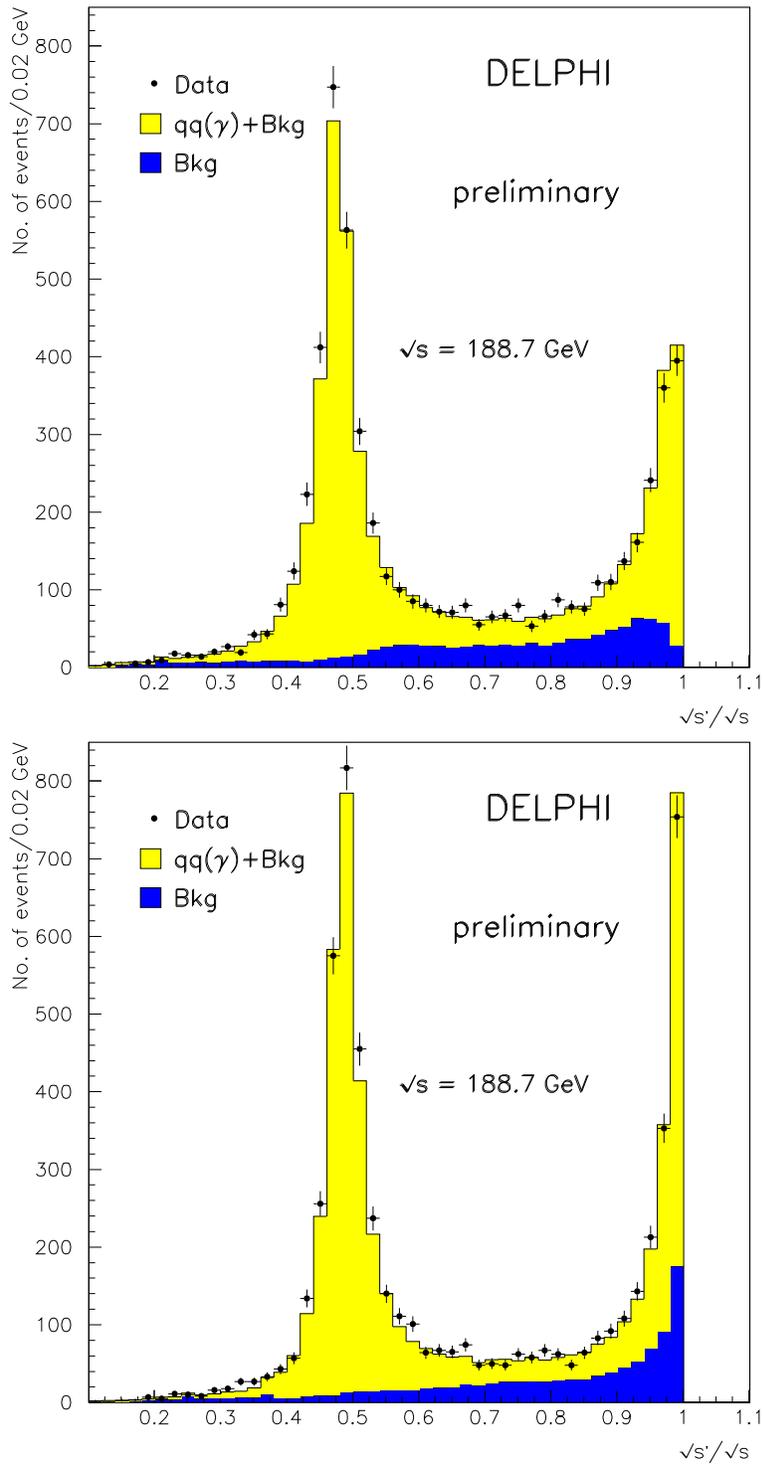

\begin{center}
\begin{tabular}{cc}
\epsfig{file=sprime_plots_ecms_188.epsi,width=10cm} \\
\epsfig{file=sprime_plus_plots_ecms_188.epsi,width=10cm}
\end{tabular}
\caption{\label{sprfig9}
SPRIME (top) and SPRIME+ (bottom) results for the data collected during 
early 1998 at a c.m. energy 
around 188~GeV (black bullets), compared to the simulated events at 
the same energy using PYTHIA. The data 
correspond to an integrated luminosity of about 54~pb$^{-1}$.} 
\end{center}
\end{figure}

The estimated effective c.m.
energy ($\sqrt{s^\prime}$), for hadronic events (selected using 
the standard DELPHI hadronic criteria, see~\cite{delphihad}) collected 
from the run around 188.7~GeV during 1998,
is shown in the upper plot of
figure~\ref{sprfig9} (data points) and is compared
to a sample simulated with PYTHIA and followed through the
complete simulation of the detector, DELSIM~\cite{perfo}.
The results show a satisfactory description of the data by
the simulation. 

High efficiencies and purities are obtained for
selecting ``high energy'' events (defined as those with effective 
energy above $0.9\sqrt{s}$), and for selecting Z radiative
return events. In the following, we shall call ``low energy'' events
those with effective 
energy below $0.9\sqrt{s}$.

\begin{figure}
\begin{center}
\begin{minipage}{0.8\linewidth}
\mbox{\epsfig{file=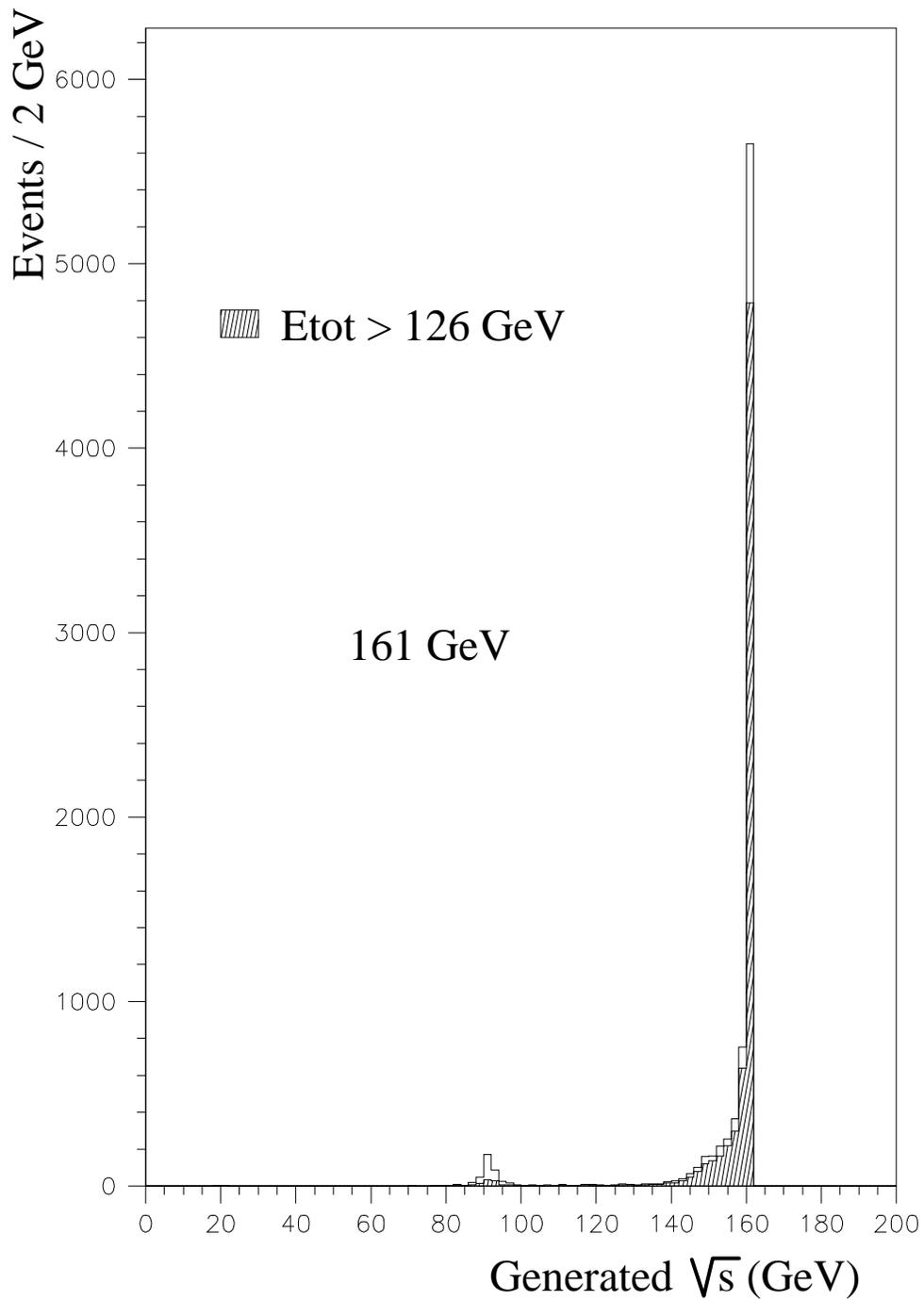,width=\linewidth}}
\caption{\label{sprfig5}
Generated effective centre of mass energy for simulated events with a 
SPRIME result above $0.9\sqrt{s}$ (solid line), and the same distribution
for events with a SPRIME result above $0.9\sqrt{s}$ and requiring that the
total energy seen in the event 
is above 126~GeV.}
\end{minipage}
\end{center}
\end{figure}

The generated effective
c.m. energy for simulated events at beam energy of 80.5~GeV, 
selected requiring a SPRIME result above $0.9\sqrt{s}$, is shown
in figure~\ref{sprfig5} (hollow histogram). The
hatched histogram shows the same
distribution for a subsample of those 
events in which the energy seen in the detector
is above $M_Z+(\sqrt{s}-M_Z)/2 = 126$ GeV.
With this cut the small contamination due to events 
with effective c.m. energy near the Z mass is
drastically reduced.

The efficiency to select high energy events 
was computed to be 88\%, with a purity of
the selected sample of 92\%. 
When a larger purity is required, a cut on 
the  total 
energy of the event may be applied. For the cut illustrated 
in the previous paragraph the purity was
increased to 96\%, with an efficiency of 74\%.

\begin{figure}
\begin{center}
\begin{minipage}{0.8\linewidth}
\mbox{\epsfig{file=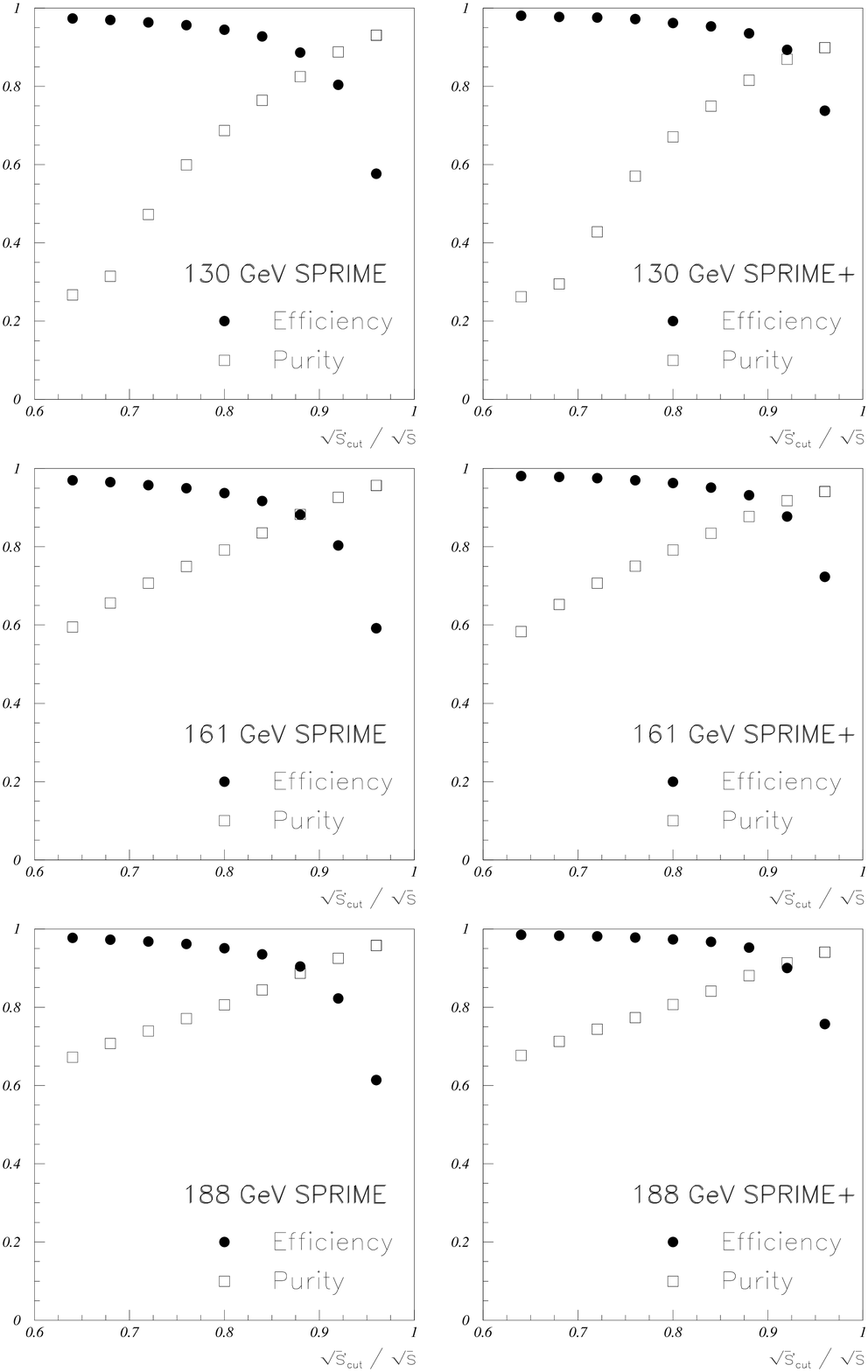,width=\linewidth}}
\caption{\label{sprfig6}
Efficiency (black bullets) and purity (white squares) for selecting events
with generated effective centre of mass energy above $0.9\sqrt{s}$ as a
function of the cut applied on the ratio of the SPRIME (left) or
SPRIME+ (right) result to the LEP c.m. energy. 
The results are shown for simulated events at 
c.m. energies of 130 GeV (two upmost plots), 161 
GeV (two central plots) and 188 GeV (two lowmost plots).}
\end{minipage}
\end{center}
\end{figure}

Using simulated events with PYTHIA at c.m. energies of 130~GeV,
161~GeV and 
188~GeV respectively, the 
efficiencies and purities for \
selecting ``high energy'' events, as a function of
the cut value in $\sqrt{s^\prime}/\sqrt{s}$, are shown in
figure~\ref{sprfig6}.

\begin{figure}
\begin{center}
\begin{minipage}{0.8\linewidth}
\mbox{\epsfig{file=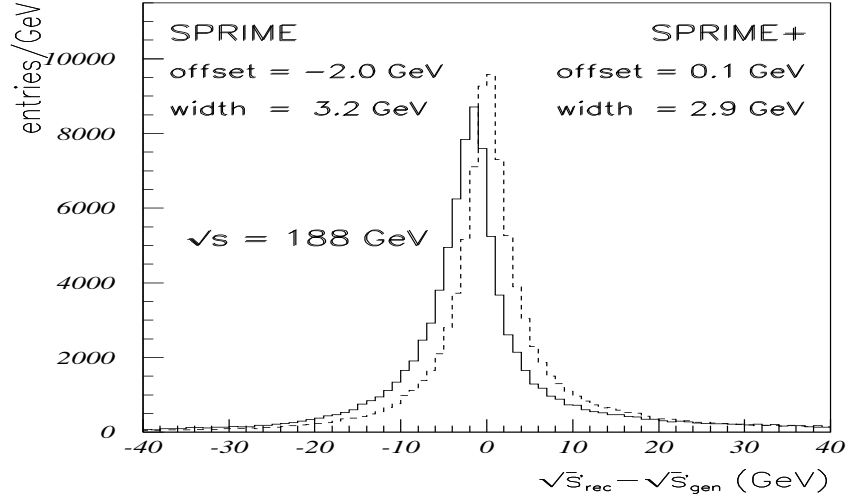,width=\linewidth,height=8cm}}
\caption{\label{sprfig7}
Solid line: 
distribution of PYTHIA simulated events at an energy in the centre of mass of 
188~GeV as a function of the difference between the SPRIME result
and the generated effective c.m. energy. Dashed line: same for SPRIME+.}
\end{minipage}
\end{center}
\end{figure}

\begin{figure}
\begin{center}
\begin{minipage}{0.8\linewidth}
\mbox{\epsfig{file=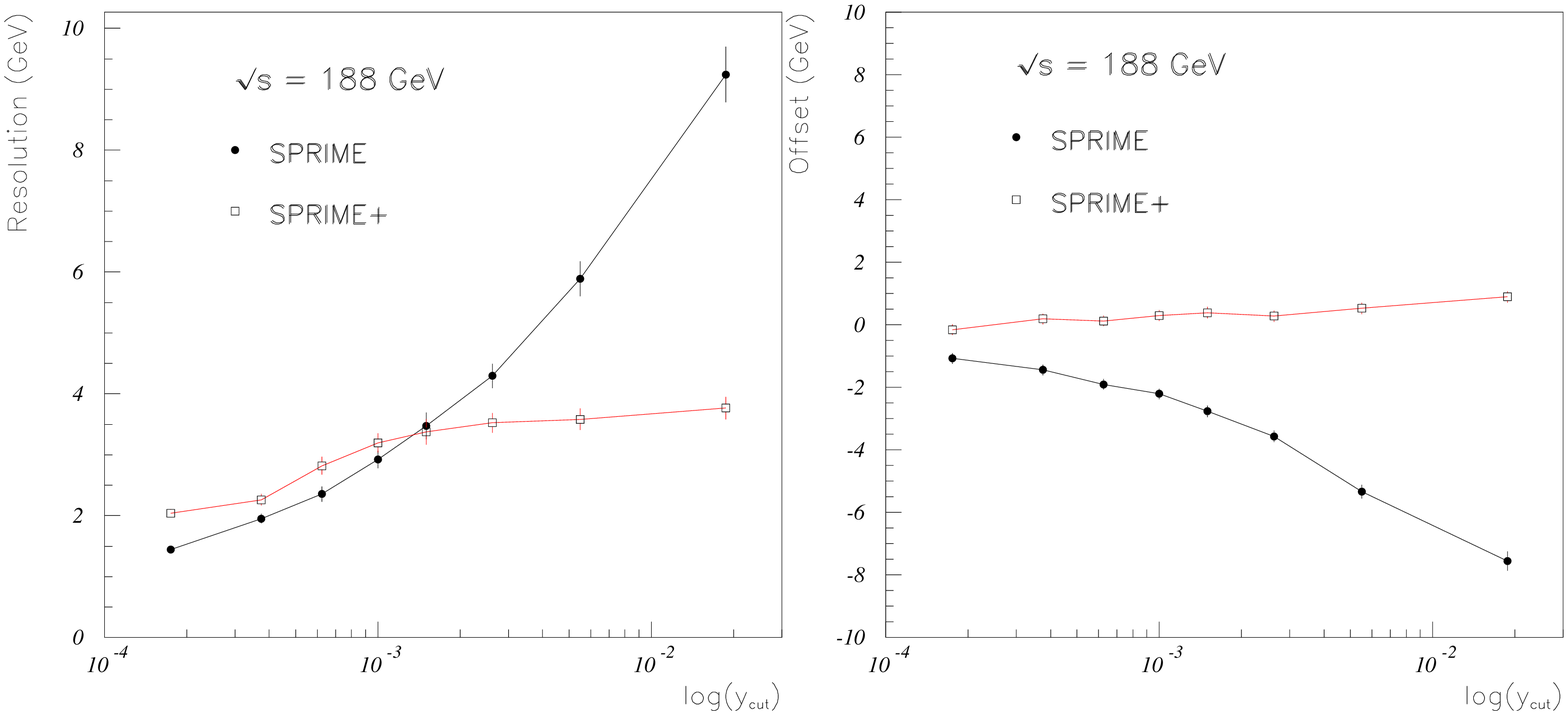,width=\linewidth,height=7cm}}
\caption{\label{sprfig8} 
Left: Resolution of SPRIME and SPRIME+, as a function of the
minimal $y_{cut}$ needed for DURHAM to reconstruct two
jets. Right: Average offset for SPRIME and SPRIME+, as a function of 
$y_{cut}$.}
\end{minipage}
\end{center}
\end{figure}

Figure~\ref{sprfig7} shows the distribution of the 
low energy events, simulated at 
c.m. energy of 188~GeV, as a function of
the difference between the SPRIME estimate and the generated effective
energy. The Full Width at Half the Maximum (FWHM)
of this distribution divided by 2.35 is used as an estimator of
the resolution of SPRIME, and is shown on the left plot of
figure~\ref{sprfig8} as a function of the
DURHAM $y_{cut}$ value needed to distinguish 2-jet from 3-jet events.
The right plot of  
figure~\ref{sprfig8} shows the average offset
of the method, defined as the peak value  of the distribution of
the difference between the SPRIME estimate and the generated effective
energy. 

\subsection{Method 2: determination of the energy of ISR photon(s)
 from a constrained fit} \indent

The second method (referred to inside the collaboration as SPRIME+)
uses a constrained fit taking into account the
4-momenta of the measured jets and photons and assuming one additional 
photon inside the beampipe. This method is able to include any number
of jets and photons and is therefore 
more appropriate than the previous method in events
with more than two jets. The fit also returns an
estimated error on the effective c.m. energy, and a
$\chi^2$ indicating the goodness of the fit. 


All particles that have not been identified as ISR photons are 
clustered into a ``natural'' number of jets using the DURHAM
jet algorithm
with a $y_{\rm cut}$ fixed to $0.002$. This $y_{\rm cut}$ value
was chosen to optimize the energy resolution; at this value,
PYTHIA reproduces the data well.

Then two kinematical constrained fits are performed, corresponding to
the hypotheses that an undetected photon went along the positive or
negative z-direction (along the beam). There are two constraints from
the transverse momentum balance,
$\sum{p_x}  =  0$ and 
$\sum{p_y}  =  0$,
and one from conservation of the total energy and of the momentum
along the beam direction:
\begin{center}
  $\sum{E} = \sqrt{s} - \sum{p_z}~~~$  or  
  $~~~\sum{E}  =  \sqrt{s} + \sum{p_z}$ 
\end{center}
assuming $\sum{p_z} > 0$ or $\sum{p_z} < 0$ respectively. 

The fitted jet momentum, $\vec{p_j}^f$, was projected onto a set of axes with
one component parallel to the measured jet momentum,
$\vec{p_j}^m$, and two transverse components.
The parallel component was described by a rescaling factor, $\exp(a_j)$,
while the transverse components were described by
parameters multiplying perpendicular momenta fixed to 1~GeV/$c$:
\begin{displaymath}
\vec{p_j}^f = \exp(a_j) \vec{p_j}^m + b_j \vec{p_j}^b
+ c_j \vec{p_j}^c.
\end{displaymath} 
To determine $\vec{p_j}^b$ and $\vec{p_j}^c$,
the jet energy, ${E_j}^m$, and thereby
also to a good approximation the jet mass, was rescaled with
the same factor $\exp(a_j)$ as the jet momentum;
the fitting algorithm then minimized a $\chi^2$:
\begin{displaymath}
\chi^2 = \sum_{j} \frac{(a_j-a_0)^2}{\sigma_{a_j}^2}
+\frac{b_j^2}{\sigma_{b_j}^2}
+\frac{c_j^2}{\sigma_{c_j}^2},
\end{displaymath}
while forcing the fitted event to obey the constraints.

The errors were
parametrized as a function
of the polar angle ${\theta}_j$ of the jet with respect to the beam:
\begin{eqnarray} a_0 & = & 0.15 + 0.40 \cdot {\rm cos}^4 {\theta}_j \nonumber
  \\ {\sigma}_{a_j} & = & 0.15 + 0.40 \cdot {\rm cos}^4 {\theta}_j
 \nonumber \\ {\sigma}_{b_j} =  {\sigma}_{c_j}  & = & 1.8 + 1.08 \cdot
 {\rm
 cos}^4 {\theta}_j ~~ {\rm GeV}\nonumber
 \end{eqnarray}
Photons are treated in the same way as jets, but with smaller
estimated measurement errors:
\begin{eqnarray} a_0 & = & 0 \nonumber
  \\ {\sigma}_{a_j} & = & 0.04  \nonumber\\ 
  {\sigma}_{b_j}  =  {\sigma}_{c_j}  & = & 0.01 ~~ {\rm GeV} \nonumber \, .
\end{eqnarray}

When, after
the fit, $\sum{p_z}$ has the same sign as assumed in the constraints,
the solution is said to be ``physical'', otherwise it is called 
``unphysical''.
The solution is chosen that has the smallest $\chi^2$ and is
physical. If both solutions are unphysical the 
measured energy and momentum balance indicate that there was no
significant photon lost inside the beampipe. In that case a new
constrained fit is performed, using the four constraints of momentum
and energy conservation with $\sum{p_z}$ fixed to zero.

The fitting algorithm used is 
similar to the method described in detail in~\cite{Wmass}.

In case the standard fit as described above does not converge, a
special procedure is applied to try to recuperate the event. 
The longitudinal errors are relaxed for the measured jets:
\begin{eqnarray} a_0 & = & 0.75 + 2.0 \cdot {\rm cos}^4 {\theta}_j \nonumber
 \\ 
 {\sigma}_{a_j} & = & 0.75 + 2.0 \cdot {\rm cos}^4 {\theta}_j \nonumber
\end{eqnarray} 
and for the photons:
\begin{eqnarray} a_0 & = & 0 \nonumber\\ 
  {\sigma}_{a_j} & = & 0.057. \nonumber
\end{eqnarray}
Then a series of fits is done with the 4 constraints of
energy and momentum conservation after assuming an additional photon 
along $z$, with
momentum $p_z$ = -100, -95, -90 ... +95, +100 GeV along the beam. 

The fit with the smallest $\chi^2$ is chosen. If there are at least
three fits around the minimum that converged, interpolation with a
parabola is used to improve the determination of the $p_z$ with the lowest
$\chi^2$. A final fit is done with $p_z$ constrained to this value. 

\subsubsection{Results}

The performance of the constrained fit method was studied using PYTHIA 
q{$\overline{\mbox{q}}\gamma$} simulation at c.m. 
energies of 130 GeV, 161 GeV and 188 GeV, 
including the full detector simulation DELSIM.

The constrained fit turns out to be quite robust. The standard fit does 
not converge in 0.4\% of the events, in which case the
recuperation method is used. This method is able to recuperate 60\%
of the events that failed. 

The resolution obtained 
with SPRIME+ can be seen from Figure \ref{sprfig7}. The
improvement in resolution and offset 
with respect to SPRIME is evident especially for
events with a high $y_{cut}$ from 3 to 2 jets, for which the assumption
of 2 massless jets is no longer valid (at a c.m. energy of 161 GeV,
46\% of the events have $y_{cut} > 10^{-3}$). 
This is shown for the
reconstruction of Z radiative return events in figure~\ref{sprfig8}. 

The improvement in resolution leads to higher
efficiencies for selecting events with
$\sqrt{s'}$ larger than $0.9\sqrt{s}$ 
(figure~\ref{sprfig6}). For very high purities, however, 
there is hardly any improvement. This is due to a side effect
of the improved resolution for 3-jet like configurations: when an
energetic ISR photon is radiated inside the detector acceptance but
not tagged as an ISR photon, it is be treated as a third jet in
SPRIME+, giving an estimated effective c.m. energy larger
than 0.9$\sqrt{s}$. As SPRIME has a worse resolution for 3-jet like
events, a significantly larger part of events of this type of events miss
the cut, resulting in a higher purity. 
 
With a cut requiring SPRIME+ larger than $0.9\sqrt{s}$, the efficiency
and purity for events with generated $\sqrt{s'}$ 
larger $0.9\sqrt{s}$ for the
PYTHIA MC sample at 161 GeV are 94\% and 90\% respectively. 

In figure~\ref{sprfig9} the data taken at 189 GeV is compared to 
simulation. The high energy peak is well seperated from the radiative 
return peak. The figure
also shows the improved resolution for the background which 
mainly consists of W pair events. The multi-jet events coming from the
fully hadronic decay are reconstructed very well at the full c.m. energy.
The large tail at low $\sqrt{s}$ consists of hadronic-leptonic
and fully leptonic W pair events containing neutrinos that are not
included in the reconstruction hypothesis.

\section{Summary}
The methods used by 
DELPHI for the determination of the effective center-of-mass 
energy in hadronic events at LEP 2 were presented.

The ``high energy'' peak is
very well separated from the radiative return to the Z, allowing for
efficiencies and purities well above 80\% to be obtained for selecting 
events with effective energy greater than $0.9\sqrt{s}$ (``high
energy'' events) or in the region of the Z mass (``Z radiative return''
events). 

Typical resolutions 
ranging from  2 to 3 GeV on the effective center-of-mass energy 
are achieved, depending on the event topology.

\subsection*{Acknowledgements}

We would like to thank the colleagues from the DELPHI collaboration, who
tested the algorithms presented in this paper helping to improve them.
We thank in particular W. Venus for reading 
and commenting the manuscript.


\newpage

\end{document}